\documentclass[preprint]{revtex4}
\usepackage{amsfonts}
\usepackage{amssymb,epsfig}
\usepackage{latexsym}
\usepackage{amsmath}
\usepackage{graphicx}

\begin{document}

\title{Power-law entropy corrected new holographic scalar field models of
dark energy with modified IR-cutoff}
\author{A.Khodam-Mohammadi\thanks{%
Email:\text{khodam@basu.ac.ir}}}
\affiliation{Department of Physics, Faculty of Science, Bu-Ali Sina University, Hamedan
65178, Iran\\
Research Institute for Astronomy and Astrophysics of Maragha
(RIAAM)-Maragha, Iran, P. O. Box: 55134-441}

\begin{abstract}
In this work, the PLECHDE model with Granda-Oliveros (G-O) IR-cutoff
is studied. The evolution of dark energy density, deceleration and
EoS parameters are calculated. I demonstrate that under a condition,
our universe can accelerate near the phantom barrier at present
time. We calculate these parameters also in PLECHDE at Ricci scale,
when when $\alpha=2$ and $\beta=1$, and at last a comparison between
Ricci scale, G-O cutoff and non-corrected HDE without matter field
with G-O cutoff is done. The correspondence between this model and
some scalar field of dark energy models is established. By this
method, the evolutionary treatment of kinetic energy and potential
for quintessence, tachyon, K-essence and dilaton fields, are
obtained. I show that the results has a good compatibility with
previous work in the limiting case of flat, dark
dominated and non corrected holographic dark energy.\\

\end{abstract}

\maketitle


\newpage

\section{Introduction}

Now a days, dark energy is one of the well known scenarios in modern
cosmology. Recent astrophysical observations such as data from distant SNIa,
LSS and CMBR, reveal that our universe treats under an accelerated expansion
\cite{Perl}. This expansion may be driven by a mysterious energy component
with negative pressure, so called, dark energy (DE), which fills $\sim3/4$
of total energy contents of our universe with an effective equation of state
(EoS) parameter $-1.48<w_{eff}<-0.72$ \cite{Hannestad}. Despite of many
efforts in this subject, the nature of DE is the most mysterious problem in
cosmology. The cosmological constant is the simplest candidate for DE,
called $\Lambda$CDM model. It has a constant energy density and pressure
with a constant equation of state. The $\Lambda$CDM model suffers two known
difficulty as follows: cosmic coincidence and fine tuning problems. The
cosmic coincidence problem requires that our universe behaves in such a form
that the ratio of dark matter to dark energy densities must be a constant of
order unity or varies more slowly than the scale factor and finally reaches
to a constant of order unity \cite{28sheykhi,Bisabr,khodam5}. In order to
avoid these difficulties, the cosmologists proposed dynamical models of DE.
The holographic dark energy (HDE) proposal, based on holographic principle,
is one of the most attractive candidates of dynamical DE, which has been
widely extended in many literatures \cite{holog}. According to the
holographic principle, the number of degrees of freedom in a bounded system
should be finite and has a relationship with the area of its boundary \cite%
{Hoft}. The holographic principle is a fundamental principle in quantum
gravity. In quantum field theory, a short distance (UV) cutoff, $\Lambda$,
is related to the long distance (infrared) cutoff, $L$, due to the limit set
by forming a black hole. In the other words, the zero-point energy of a
system with size $L$ should not exceed the mass of a black hole with the
same size. This fact directs us to $L^3\Lambda^3\leq(M_{P}L)^{3/2}$ \cite%
{8,coh99}, where $M_{P}$ is reduced Plank mass. From this inequality, one
can obtain a limit for energy density corresponding to the zero point energy
and cutoff $\Lambda$ as $\rho_{\Lambda}\leq M_{P}^2L^{-2}$ or $%
\rho_{\Lambda}=3n^2M_{P}^2L^{-2}$, where $\rho_{\Lambda}\sim\Lambda^4$. Here
$n$ is a numerical constant and coefficient $3$ is given for convenience
\cite{li1,li04}. The IR-cutoff $L$ plays an essential role in this model. If
$L$ is chosen as particle horizon, the HDE can not drive an acceleration
expansion \cite{hsu}, while for future event horizon, Hubble scale ``$%
L=H^{-1}$", and apparent horizon as an IR-cutoff (AH-IR-cutoff), an
accelerated expansion can be driven by HDE model and the coincidence problem
can also be solved \cite{Pavon1,28sheykhi,Sheykhi1, odintsov}. More
recently, a model of interacting HDE at Ricci's scale, in which $L=(\dot{H}%
+2H^2)^{-1/2}$ has been proposed. They performed an extending discussion on
the cosmic coincidence problem, age problem and obtained some observational
constraints on their's model \cite{Pavon2}. Granda and Oliveros proposed a
new IR-cutoff for holographic DE (named new holographic DE) which include a
term proportional to $\dot{H}$ \citep{Granda2, granda1}. Despite of the
holographic dark energy based on the event horizon, this model depends on
local quantities, avoiding in this way the causality problem. They showed
that power law expansion can appear as the solution of friedmann equations.
Their model can generate scalar field potentials which give rise to scaling
solutions in a FRW cosmological background. Also the author with
collaborators, studied the Cosmological evolution and statefinder diagnostic
for new holographic DE model in non flat universe \cite{newkhodam}.

Although it has not been proposed any well quantum field theory prescription
of DE scenario, it is believed that the thermodynamical description of an
accelerating universe may reveal the nature of it. In the HDE model, the
area law of entropy, ``$S_{\mathrm{BH}} = A/(4G)$", is satisfied on the
horizon \cite{Wald}. Here $A\sim L^2$ is the area of horizon. Therefore this
model is strongly connected to entropy of spacetime in Einstein gravity. Any
correction to entropy, affects directly on the energy density of HDE.

The entropy-corrected dark energy models based on quantum field theory and
gravitation have been widely extended by many authors in the recent years
\cite{Pavon5,E-C}. The motivation of these corrections has been based on
black hole physics, where some gravitational fluctuations and field
anomalies can affect the entropy-area law of black holes. The logarithmic
corrections and power-law corrections are two procedure in dealing with this
fluctuations. We know that the gravitation is the base of cosmology. The
gravitational entropy plays a crucial role in this connection.

The ``power-law corrected entropy (PLEC)" is appeared in dealing with the
entanglement of quantum fields in and out of the horizon \cite{Das2}. The
entropy of PLEC is given by \cite{Radicella}
\begin{equation}
S=\frac{A}{4G}[1-K_{\gamma}A^{1-\gamma/2}],  \label{entpl}
\end{equation}
where $\alpha$ is a dimensionless positive constant and
\begin{equation}
K_{\gamma}=\frac{\gamma}{4-\gamma}(4\pi r_{c}^2)^{\gamma/2-1}.
\label{kalpha}
\end{equation}
Here $r_{c}$ is the crossover scale. Further details are referred by \cite%
{Das2,Radicella,Das3}. It is worthwhile to mention that in the most
acceptable range of $4>\gamma>2$ \cite{Das2,Radicella}, the correction term
(i.e. the second term of (\ref{entpl})), is effective only at small $A$'s
and it falls off rapidly in large values of $A$. Therefore, by large horizon
area, the entropy-area law is recovered. However the thermodynamical
considerations predict that the case $\gamma\leq2$ may be acceptable, but it
should be removed for cosmic coincidence consideration \cite{khodam5}. Due
to entropy corrections to the Bekenstein-Hawking entropy ($S_{BH}$), the
Friedmann equation should be modified \cite{Pavon5}. In comparison with
ordinary Friedman equation, the energy density of PLECHDE, has been given by
\cite{Sheykhi2}
\begin{equation}
\rho_{D}=3n^2M_{p}^{2}L^{-2}-\delta M_{p}^{2}L^{-\gamma},  \label{endpl}
\end{equation}
where $\delta$ and $\gamma$ are the parameters of PLECHDE model. The
ordinary HDE is recovered for $\delta=0$ or $\gamma=2$.

Recently, the HDE and agegraphic/newagegraphic dark energy (ADE/NADE) models
have been extended regarding the entropy corrections (ECHDE, PLECHDE,
PLECNADE) and a reconstruction with $F(R)$ gravity has been performed \cite%
{Karami2}.

The outline of my paper is as follows: In Sec. \ref{PLECHDE}, the PLECHDE
model with G-O IR-cutoff is studied and the evolution of dark energy,
deceleration parameter and EoS parameter are calculated. In Sec. \ref{scalar}%
, the correspondence between this model and some scalar field of dark energy
models is established. The paper is finished with some concluding remarks.

\section{General formalism of PLECNHDE model}

\label{PLECHDE} The energy density of PLECHDE model with G-O IR-cutoff in
Planck mass unit, in which $(8\pi G)^{-1/2}=M_{P}=1$, can be given by
\begin{equation}
\rho_{D}=3(\alpha H^{2}+\beta\dot{H})-\delta (\alpha H^{2}+\beta\dot{H})^{%
\frac{\gamma}{2}},  \label{rpl}
\end{equation}
where we are using G-O scale as: $L_{GO}=(\alpha H^{2}+\beta\dot{H})^{-1/2}$
\cite{granda1}, including two constants, $\alpha$ and $\beta$, with the
Hubble parameter $H$ and its time derivative $\dot{H}$. Here also $\delta$
and $\gamma$ are two parameters of the PLECHDE model. The line element
metric of a non-flat FRW universe with spacial curvature parameter $k$ is
\begin{eqnarray}
ds^2=dt^2-a^2(t)\left(\frac{dr^2}{1-kr^2}+r^2d\theta^2+r^2sin(\theta)^2d%
\phi^2\right),  \label{metric}
\end{eqnarray}
where $a(t)$ is the scale factor, and $k = -1, 0, 1$ corresponds to the
open, flat, and closed universe, respectively. The Friedmann equation for a
non-flat universe dominated with two dark components, energy and matter, is
written by
\begin{eqnarray}
H^2+\frac{k}{a^2}=\frac{1}{3}(\rho_{D}+\rho_{m}).  \label{Fried}
\end{eqnarray}
By introducing, as usual, the fractional energy densities as
\begin{equation}
\Omega_{m}=\frac{\rho_{m}}{\rho_{cr}},~~\Omega_{K}=\frac{K}{H^2a^2}%
,~~\Omega_{D}=\frac{\rho_{D}}{\rho_{cr}}=L_{GO}^{-2}H^{-2}\left(1-\frac{%
\delta}{3}L_{GO}^{-\gamma+2}\right),  \label{fed}
\end{equation}
where $\rho_{cr}=3H^2$ is the critical energy density, the Friedmann
equation will be rewritten as
\begin{equation}
\Omega_{D}+\Omega_{m}=1+\Omega_{K}.  \label{Freq2}
\end{equation}
Once again to preserve the Bianchi identity or local energy-momentum
conservation law, $\nabla_{\mu}T^{\mu\nu}=0$, the total energy density $%
\rho_{tot} =\rho _{D}+\rho _{m}$ is satisfied in the following equation
\begin{equation}
\dot{\rho_{tot}}+3H(1+w)\rho_{tot} =0  \label{CE0}
\end{equation}%
where $w=p_{tot}/\rho_{tot} $ is the total equation of state (EoS)
parameter. In the absence of any interaction between dark energy and
pressureless cold dark matter (CDM) with subscript '$m$', two energy
densities $\rho _{D}$ and $\rho _{m}$ are conserved separately and the
conservation equations are written as
\begin{eqnarray}
\dot{\rho}_{D}+3H(1+w_{D})\rho _{D} &=&0,  \label{CE1} \\
\dot{\rho}_{m}+3H\rho _{m} &=&0.  \label{CE2}
\end{eqnarray}%
Using Eq. (\ref{rpl}) and the time derivative of G-O scale: $\dot{L_{GO}}%
=-H^3L^{3}_{GO}(\alpha \dot{H}/H^2+\beta \ddot{H}/2H^3)$, the time
derivative of $\rho_{D}$ is
\begin{equation}
\dot{\rho_{D}}=6H^{3}\left(\alpha \frac{\dot{H}}{H^2}+\beta \frac{\ddot{H}}{%
2H^3}\right)\left(1-\frac{\gamma\delta}{6} L_{GO}^{-\gamma+2}\right) .
\label{rod}
\end{equation}
Taking time differential of Eq. (\ref{Fried}) and using Eqs. (\ref{rod}, \ref%
{fed}, \ref{Freq2}, \ref{CE2}), we find
\begin{equation}
\alpha \frac{\dot{H}}{H^2}+\beta \frac{\ddot{H}}{2H^3}=\frac{\left[1+\frac{%
\dot{H}}{H^2}-(\frac{u}{2}-1)\Omega_{D}\right]}{1-\frac{\gamma\delta}{6}
L_{GO}^{-\gamma+2}} .  \label{eq10}
\end{equation}
Also from G-O scale and $\Omega_{D}$ in (\ref{fed}), we have, $\dot{H}%
/H^2=\left(\frac{\Omega_{D}}{1-\frac{\delta}{3}L_{GO}^{-\gamma +2}}%
-\alpha\right)/\beta$, therefore the Eq. (\ref{rod}) yields
\begin{equation}
\dot{\rho_{D}}=3H^3\left[\frac{2}{\beta}\left(\frac{\Omega_{D}}{1-\frac{%
\delta}{3}L_{GO}^{-\gamma +2}}-\alpha+\beta\right)+(u-2)\Omega_{D})\right] ,
\label{rod2}
\end{equation}
where $u=\rho _{m}/\rho _{D}=\Omega_m/\Omega_D$ is the ratio of energy
densities.

Differentiating of $\Omega_{D}$ with respect to cosmic time and using $%
\dot\Omega_{D}=H\Omega_{D}^{\prime}$, gives
\begin{equation}
\Omega_{D}^{\prime}=\left[\frac{2}{\beta}\left(\frac{\Omega_{D}}{1-\frac{%
\delta}{3}L_{GO}^{-\gamma +2}}-\alpha+\beta\right)\right]+u\Omega_{D},
\label{rod3}
\end{equation}
where the dot and prime denote the derivative with respect to the cosmic
time and the derivative with respect to $x=ln~a$, respectively. At last,
using Eqs. (\ref{fed}, \ref{CE1}, \ref{rod2}), the EOS parameter and
deceleration parameter $q=-1-\dot{H}/H^2$ as a function of $%
L_{GO},~\Omega_{D}$ and $H$, can be obtained as
\begin{eqnarray}
&&w_{D}=-\frac{2}{3\beta\Omega_{D}}\left(\frac{\Omega_{D}}{1-\frac{\delta}{3}%
L_{GO}^{2-\gamma}}+\beta-\alpha\right)-\frac{1}{3}(1+u),  \label{Eos1} \\
&&q=\frac{1}{\beta}\left(\alpha-\beta-\frac{\Omega_{D}}{1-\frac{\delta}{3}%
L_{GO}^{2-\gamma}}\right) ,  \label{q1}
\end{eqnarray}
where from (\ref{fed}) and (\ref{q1}), we have
\begin{equation}
\frac{\Omega_{D}}{1-\frac{\delta}{3}L_{GO}^{2-\gamma}}%
=L_{GO}^{-2}H_{GO}^{-2}=\alpha-\beta-\beta q.  \label{eq21}
\end{equation}
At Ricci scale where $\alpha=2$ and $\beta=1$, these parameters yield
\begin{eqnarray}
&&w^{R}_{D}=-\frac{2}{3\Omega_{D}}\left(\frac{\Omega_{D}}{1-\frac{\delta}{3}%
L_{R}^{2-\gamma}}+-1\right)-\frac{1}{3}(1+u),  \label{Eos2} \\
&&q^R=1-\frac{\Omega_{D}}{1-\frac{\delta}{3}L_{R}^{2-\gamma}} ,
\label{q2}
\end{eqnarray}
find Also from Eqs. (\ref{CE1}, \ref{CE2}), the evolution of $u$, is
governed by
\begin{equation}
\dot{u}=3Huw_{D}.  \label{INTQ}
\end{equation}
At present time ($\Omega_{D}\approx 2/3,~u\approx 0.4$) from (\ref{Eos1}, %
\ref{q1}), the EoS parameter become: $w_{D}\approx q-0.47$. The universe
exist in accelerating phase ($q<0$) if $w_{D}< -0.47$ and the phantom
divide, $w_{D}\leqslant-1$, may be crossed provided that $q\lesssim -0.53$.
This condition give us: $\dot{H_{0}}/H_{0}^2\gtrsim -0.47$ and from Eq. (\ref%
{q1}), we have $L_{GO-0}^{-2}H_{0}^{-2}\gtrsim \alpha-0.47\beta$. Also the
transition between deceleration ($q>0$) to acceleration ($q<0$) phases took
place at $L_{GO-acc}^{-2}H_{acc}^{-2}=\alpha-\beta$. However from recent
analysis of SNe+CMB data with the $\lambda$CDM model, our universe began to
accelerate at redshift around $z\sim 0.52-0.73$ \cite{Perl}.

Recently, Wang and Xu \cite{Wang1}, by using some observational data, have
constrained the new HDE model in non flat universe. The best fit values of
the new HDE model parameters ($\alpha,~\beta$) with their confidence level
are obtained as: $\alpha=0.8824^{+0.2180}_{-0.1163}~(1%
\sigma)~^{+0.2213}_{-0.1378}~(2\sigma)$, $%
\beta=0.5016^{+0.0973}_{-0.0871}~(1\sigma)~^{+0.1247}_{-0.1102}~(2\sigma)$.
Therefore the PLECNHDE crossing the phantom barrier at present time
($z=0$), if: $L_{GO-0}^{-2}H_{0}^{-2}\gtrsim 0.65$. At
deceleration acceleration phase transition, we find: $%
L_{GO-acc}^{-2}H_{acc}^{-2}\sim 0.38$. It is worthwhile to mention that at
Ricci scale, at present time, the dark energy behaves as phantom if $%
L_{GO-0}^{-2}H_{0}^{-2}> 1.53$ and our universe started the acceleration
expansion from $L_{GO-acc}^{-2}H_{acc}^{-2}=1$.

In limiting case for ordinary new holographic ($\delta=0$) in flat universe
without any matter field, ($u=0$ , $\Omega_{D}=L_{GO}^{-2}H^{-2}=1$), we
find
\begin{equation}
H=\frac{\beta}{\alpha-1}\frac{1}{t},  \label{HO}
\end{equation}
and Eqs. (\ref{Eos1}, \ref{q1}) reach to
\begin{eqnarray}
&&w^{O}_{D}=\frac{2}{3}\frac{\alpha-1}{\beta}-1, \\
&&q^{O}=\frac{\alpha-1}{\beta}-1.
\end{eqnarray}
where have a good compatibility with \cite{granda1}.

In table (\ref{tb1}), we compare three cases: (\textit{I}). non
corrected HDE with G-O cutoff, $u=0$ and $\Omega_{D}=1$, (NHDE),
(\textit{II}). PLECHDE with G-O cutoff (PLECHDE-GO) and
(\textit{III}). PLECHDE in Ricci scale (PLECHDE-R).

\begin{table*}[tbp]
\caption{Comparison of $L^{-2}H^{-2}$, $q$, and $w$ in three models of DE}%
\begin{tabular}{|c|c|c|c|c|}
\hline DE models & $L_{acc}^{-2}H_{acc}^{-2}$ &
L$_{0}^{-2}$H$_{0}^{-2}$ & $q_0$ & $w_0$
\\ \hline NHDE & 1 & 1 & -1.23 & -1.56 \\ \hline PLECHDE-R & 1 & 1.53 & -0.53 & $\sim$ -1\\ \hline
PLECHDE-GO & 0.38 & 0.65 & -0.54 & $\sim$ -1\\ \hline
\end{tabular}%
\label{tb1}
\end{table*}
In this table we see that our universe may be behave around phantom
barrier at present time if we use PLECHDE as dark energy model with
G-O cutoff or Ricci scale. In the case NHDE, the evolution of our
universe can express only at phantom phase, far from the phantom
barrier. Recent observational data suggest that $w$ does not depart
from $-1$ at sufficiently low redshift or present time \cite{Serra}.
\section{PLECNHDE scalar field models}

\label{scalar} In this section the correspondence between PLECNHDE and
famous scalar field models of DE such as quintessence, tachyon, K-essence
and dilaton are established. We can do this correspondence by comparing the
PLECHDE density with the energy density of the scalar field model and also
equating their EoS parameters. At last the dynamics of scalar field and its
potential for various scalar fields, are obtained.

\subsection{PLECHDE quintessence model}

The energy density and pressure density of the quintessence scalar field are
given by \cite{Edmund}
\begin{equation}
\rho_{\phi}=\frac{1}{2}\dot\phi ^2+V(\phi),~~~~p_{\phi}=\frac{1}{2}\dot\phi
^2-V(\phi).
\end{equation}
The EoS parameter for scalar field is given by
\begin{equation}
w_{\phi}=\frac{\dot\phi ^2-2V(\phi)}{\dot\phi ^2+2V(\phi)}.  \label{wp1}
\end{equation}
After comparing the EoS parameters of PLECNHDE (\ref{Eos1}) with scalar
field (\ref{wp1}), and equating the corresponding energy densities, we find
\begin{eqnarray}
&&\frac{1}{2}\dot\phi ^2+V(\phi)=3L_{GO}^{-2}-\delta L_{GO}^{-\gamma} \\
&&\frac{\dot\phi ^2-2V(\phi)}{\dot\phi ^2+2V(\phi)}=-\frac{2}{%
3\beta\Omega_{D}}\left(\frac{\Omega_{D}}{1-\frac{\delta}{3}L_{GO}^{2-\gamma}}%
+\beta-\alpha\right)-\frac{1}{3}(1+u).
\end{eqnarray}
By solving these equations, the dynamics of scalar field and the potential
are obtained as
\begin{eqnarray}
\phi^{\prime 2}&=&\frac{2}{\beta }\left[\left(\alpha-\beta-\frac{\Omega_{D}}{%
1-\frac{\delta}{3}L_{GO}^{2-\gamma}}\right)+\beta \Omega_{D}(1-\frac{u}{2})%
\right],  \label{eq12} \\
V(\phi)&=&\frac{H^2}{\beta }\left[\left(\frac{\Omega_{D}}{1-\frac{\delta}{3}%
L_{GO}^{2-\gamma}}+\beta-\alpha\right)+\beta\Omega_{D}(\frac{u}{2}+2)\right]
.  \label{eq13}
\end{eqnarray}
Integrating Eq. (\ref{eq12}) with respect to the scale factor `$a$' yields
the evolutionary form of the quintessence scalar field as
\begin{equation}
\phi(a)-\phi(a_{0})=\int_{a_{0}}^{a}{\frac{da}{a \sqrt{\beta}}\left[%
2\left(\alpha-\beta-\frac{\Omega_{D}}{1-\frac{\delta}{3}L_{GO}^{2-\gamma}}%
\right)+\beta \Omega_{D}(2-u)\right]^{1/2}}
\end{equation}
In the limiting case, same as end of previous section, for $%
\delta=\Omega_{K}=u=0$ and $\Omega_{D}=1$, exactly as the same as \cite%
{granda1}, we obtain
\begin{eqnarray}
\phi&=&\sqrt{\frac{2\beta}{\alpha-1}}\ln t,  \label{eq14} \\
V(\phi)&=&\frac{3\beta-\alpha+1}{(\alpha-1)^2}\exp(-\sqrt{\frac{2(\alpha-1)}{%
\beta}}) .  \label{eq15}
\end{eqnarray}

\subsection{PLECHDE tachyon model}

One of the well known scalar field that has been considered as the source of
dark energy is the tachyon field \cite{29G,30G}. It is an unstable field
which can be used in string theory through its role in the Dirac-Born-Infeld
(DBI) action to describe the D-bran action \cite{7}. The effective
Lagrangian for the tachyon field is given by
\begin{equation*}
\mathcal{L}=-V(\phi )\sqrt{1-g^{\mu \nu }\partial _{\mu }\phi \partial _{\nu
}\phi },
\end{equation*}%
where $V(\phi )$ is the potential of tachyon. The energy density and
pressure of the tachyon field are given by \cite{7}
\begin{equation}
\rho _{\phi }=\frac{V(\phi )}{\sqrt{1-\dot{\phi}^{2}}},  \label{tach1}
\end{equation}%
\begin{equation}
p_{\phi }=-V(\phi )\sqrt{1-\dot{\phi}^{2}}.
\end{equation}%
The EoS parameter of tachyon can be obtained as
\begin{equation}
w_{\phi }=\frac{p_{\phi }}{\rho _{\phi }}=\dot{\phi}^{2}-1.  \label{eos_tach}
\end{equation}%
The correspondence between the interacting PLECHDE and the tachyon scalar
field model can be stablished, by comparing Eqs.(\ref{rpl}) and (\ref{tach1}%
), and equating Eqs.(\ref{Eos1}) and (\ref{eos_tach}). By performing these
actions, the dynamics of scalar field and potential are given by
\begin{eqnarray}
&\dot{\phi}^2&=1+w_{D}=\frac{2}{3\beta\Omega_{D}}\left(\alpha-\beta-\frac{%
\Omega_{D}}{1-\frac{\delta}{3}L_{GO}^{2-\gamma}}\right) +\frac{1}{3}(2-u),
\label{cor1} \\
&V(\phi)&=\rho_{D}\sqrt{-w_{D}}=H^2\sqrt{\frac{3\Omega_{D}}{\beta}}\sqrt{%
2\left(\frac{\Omega_{D}}{1-\frac{\delta}{3}L_{GO}^{2-\gamma}}%
+\beta-\alpha\right) +\beta \Omega_{D}(1+u)}.  \label{cor2}
\end{eqnarray}%
The evolutionary form of the tachyon scalar field is obtained as
\begin{equation}
\phi(a)-\phi(a_{0})=\int_{a_{0}}^{a}{\frac{da}{a H} \left[\frac{2}{%
3\beta\Omega_{D}}\left(\alpha-\beta-\frac{\Omega_{D}}{1-\frac{\delta}{3}%
L_{GO}^{2-\gamma}}\right) +\frac{1}{3}(2-u)\right]^{1/2}}.
\end{equation}
In the limiting case, for $\delta=\Omega_{K} =u=0$, the dynamics of scalar
field for a universe which is filled only by new holographic tachyon dark
energy ($\Omega_{D}=1$), can be obtained
\begin{equation}
\dot{\phi}=\sqrt{\frac{2(\alpha-1)}{3\beta}}.
\end{equation}%
After integration with respect to time, and using Eqs. (\ref{cor2}, \ref{HO}%
), one obtains
\begin{eqnarray}
&&\phi =\sqrt{\frac{2(\alpha-1)}{3\beta}}t,  \label{Nphi} \\
&&V(\phi)=\frac{2\beta}{(\alpha-1)\phi^2}\sqrt{\frac{1+3\beta-2\alpha}{3\beta%
}},  \label{NV}
\end{eqnarray}
where we assumed the integration constant equal to zero. These relations are
in exact agreement with \cite{granda1}.

\subsection{PLECHDE K-essence model\label{sec4}}

Another famous scalar field which can explain the late time acceleration of
the universe is K-essence scalar field. It has been considered by many
authors for dark energy modeling. The general scalar field action for
K-essence model as a function of $\phi $ and $\chi =\dot{\phi}^{2}/2$ is
given by \cite{41G}
\begin{equation}
S=\int d^{4}x\sqrt{-g}\text{ }p(\phi ,\chi ),
\end{equation}%
where the Lagrangian density $p(\phi ,\chi )$ relates to a pressure and
energy densities through the following relations:
\begin{eqnarray}
p(\phi ,\chi )=f(\phi )(-\chi +\chi ^{2}),  \label{pk} \\
\rho (\phi ,\chi )=f(\phi )(-\chi +3\chi ^{2}) .  \label{Rok}
\end{eqnarray}%
In this model $\dot{\phi}^{2}=2\chi $. The EoS parameter of K-essence scalar
field is given by
\begin{equation}
\omega _{K}=\frac{p(\phi ,\chi )}{\rho (\phi ,\chi )}=\frac{\chi -1}{3\chi -1%
}.  \label{w_k}
\end{equation}%
After equating the EoS parameters of K-essence field and PLECHDE, we can
find the expression for dynamics of scalar field $\dot{\phi}^2$ as
\begin{equation}
\dot{\phi}^2=2\chi=2\frac{w_{D}-1}{3w_{D}-1}=\frac{2}{3}\left[1+\frac{2\beta
\Omega_{D}}{2\left(\frac{\Omega_{D}}{1-\frac{\delta}{3}L_{GO}^{2-\gamma}}%
+\beta-\alpha\right)+\beta\Omega_{D}(u+2)}\right] .  \label{phi3}
\end{equation}
By equating Eqs.(\ref{Rok}) with (\ref{rpl}), using Eq.(\ref{eq21}), we get
the expression for $f(\phi)$ as
\begin{equation}
f(\phi)=\frac{9H^2}{2\beta}\frac{\left[2\left(L_{GO}^{-2}H^{-2}+\beta-\alpha%
\right) +\beta\Omega_{D}(u+2)\right]^2}{2\left(L_{GO}^{-2}H^{-2}+\beta-%
\alpha\right)+\beta\Omega_{D}(u+4)}.  \label{fphi}
\end{equation}%
The evolutionary form of the K-essence scalar field is obtained as
\begin{equation}
\phi(a)-\phi(a_{0})=\sqrt{\frac{2}{3}}\int_{a_{0}}^{a}{\frac{da}{a H}\left[1+%
\frac{2\beta \Omega_{D}}{2\left(\frac{\Omega_{D}}{1-\frac{\delta}{3}%
L_{GO}^{2-\gamma}}+\beta-\alpha\right)+\beta\Omega_{D}(u+2)}\right]^{1/2}}.
\end{equation}
In the limiting case, for $\delta=\Omega_{K} =u=0$ and $\Omega_{D}=1$, the
dynamics of scalar field and $f(\phi)$ reached to
\begin{eqnarray}
\dot{\phi}&=&\sqrt{\frac{2}{3}\left(\frac{3\beta-\alpha+1}{2\beta-\alpha+1}%
\right)} \\
f(\phi)&=&\frac{6\beta(2\beta-\alpha+1)}{(\alpha-1)^2\phi^2}.
\end{eqnarray}
which had been obtained by \cite{granda1}

\subsection{PLECHDE Dilaton model \label{sec5}}

The Lagrangian density (pressure) in this model is described by
\begin{equation}
p_{d}=-\chi +ce^{\lambda \phi }\chi ^{2},
\end{equation}%
where $c$ and $\lambda $ are positive constant and $\chi=\dot{\phi}^2$. The
dilaton scalar field is originated from the lowe-energy string action \cite%
{dilaton}.

The corresponding energy density is given by
\begin{equation}
\rho _{d}=-\chi +3ce^{\lambda \phi }\chi ^{2},  \label{rhod}
\end{equation}%
The correspondence between PLECHDE and dilaton scalar field gives us
\begin{eqnarray}
\omega _{d}&=&\frac{-1+ce^{\lambda \phi }\chi }{-1+3ce^{\lambda \phi }\chi }%
=-\frac{2}{3\beta\Omega_{D}}\left(\frac{\Omega_{D}}{1-\frac{\delta}{3}%
L_{GO}^{2-\gamma}}+\beta-\alpha\right)-\frac{1}{3}(1+u),  \label{dil1} \\
\rho_{d}&=&-\chi +3ce^{\lambda \phi }\chi ^{2}=3L_{GO}^{-2}-\delta
L_{GO}^{-\gamma}  \label{dilaton2}
\end{eqnarray}%
from these equations, the quantity $\chi c e^{\lambda\phi}$ is given by eq.(%
\ref{dil1})
\begin{equation}
\chi c e^{\lambda\phi}=\frac{w_{D}-1}{3w_{D}-1}=\frac{1}{3}\left[1+\frac{%
2\beta\Omega_{D}}{2\left(\frac{\Omega_{D}}{1-\frac{\delta}{3}%
L_{GO}^{2-\gamma}}+\beta-\alpha\right)+\beta\Omega_{D}(u+2)}\right].
\label{dil2}
\end{equation}
Using $\chi =\dot{\phi}^{2}/2$, one can rewrite (\ref{dil2}) with respect to
$\phi $ as
\begin{eqnarray}
&&\frac{d}{dt}e^{\lambda \phi /2} =\frac{\lambda }{\sqrt{6c}}\sqrt{1+\frac{%
2\beta\Omega_{D}}{2\left(\frac{\Omega_{D}}{1-\frac{\delta}{3}%
L_{GO}^{2-\gamma}}+\beta-\alpha\right)+\beta\Omega_{D}(u+2)}},  \notag
\end{eqnarray}
the evolutionary form of the dilaton scalar field is written as%
\begin{equation}
\phi (a) =\frac{2}{\lambda }\ln \Big{\{}e^{\lambda \phi (a_{0})/2}+\frac{%
\lambda }{\sqrt{6c}}\int_{a_{0}}^{a}\frac{d a}{a H} \sqrt{1+\frac{%
2\beta\Omega_{D}}{2\left(\frac{\Omega_{D}}{1-\frac{\delta}{3}%
L_{GO}^{2-\gamma}}+\beta-\alpha\right)+\beta\Omega_{D}(u+2)}}\Big{\}}.
\label{pdilaton}
\end{equation}%
In the limiting case, for $\delta=\Omega_{K} =u=0$ and $\Omega_{D}=1$, the
dilaton scalar field is obtained as
\begin{equation}
\phi(t)=\frac{2}{\lambda}\ln\left(\frac{\lambda}{\sqrt{6c}}\sqrt{\frac{%
1+3\beta-\alpha}{1+2\beta-\alpha}}t\right),
\end{equation}
where it has a good compatibility with \cite{granda1}

\section{conclusion}

I have been extended the work of Granda and Oliveros (G-O)
\cite{granda1} to power law entropy corrected HDE model. This model
has been arisen from the black hole entropy which may lie in the
entanglement of quantum field between inside and outside of the
horizon. The evolution of energy density, deceleration and EoS
parameter of the new PLECHDE in the context of the non-flat universe
was obtained. We shaw that in contrast with NHDE model, in this
model the evolution of our universe near the present time ($z=0$) is
compatible with recent observational data which suggest that $w\sim
-1$. In NHDE model, $w$ is far from $-1$. A comparison between NHDE,
PLECHDE with G-O cutoff and PLECHDE in Ricci scale was done. The
phantom divide can be crossed at present time if $
L_{GO-0}^{-2}H_{0}^{-2}\gtrsim 0.65;~q_0=-0.54;~w_0\sim -1$ in
PLECHDE-GO model, $L_{R-0}^{-2}H_{0}^{-2}\gtrsim
1.53;~q_0=-0.53;~w_0\sim -1$ in PLECHDE-R model. In dark dominated
flat universe without any entropy correction, we have
$L_{GO-0}^{-2}H_{0}^{-2}=1;~q_0=-1.23;~w=-1.56$. Also the transition
between deceleration to acceleration phases of expansion, took place
at $L_{GO-acc}^{-2}H_{acc}^{-2}\sim 0.38$ in PLECHDE-GO case, while
it happened at $L_{GO-acc}^{-2}H_{acc}^{-2}\sim 1$ in PLECHDE-R
model. The correspondence between PLECHDE model and some scalar
field of dark energy models has been established. The evolutionary
treatment of kinetic energy and potential for quintessence, tachyon,
K-essence and dilaton fields, are obtained. I show that the results
are in exact agreement with previous work in the limiting case of
flat, dark dominated and non corrected holographic dark energy which
is obtained by Granda and Oliveros \cite{granda1}.

\acknowledgments {This work has been supported by Research Institute
for Astronomy \& Astrophysics of Maragha (RIAAM) under research
project No. 1/2334,}

\end{document}